\documentclass[a4paper]{article}
\usepackage[english]{babel}
\usepackage[utf8x]{inputenc}
\usepackage[T1]{fontenc}

\usepackage[a4paper,top=3cm,bottom=2cm,left=3cm,right=3cm,marginparwidth=1.75cm]{geometry}

\usepackage{amsmath}
\usepackage{graphicx}
\usepackage[colorinlistoftodos]{todonotes}
\usepackage[colorlinks=true, allcolors=blue]{hyperref}

\usepackage{soul}

\title{Eigenmode beam optimisation  for optical micro-manipulation 
\footnote{This paper collates  work presented at the Bremen Workshop on Light Scattering, 2017\cite{Mazilu:2017ac},  SPIE Nanoscience + Engineering, 2017 \cite{Mazilu:2017aa}, SPIE Nanophotonics Australasia Conference, 2017\cite{Mazilu:2017ab} and SPIE Photonics West, 2019 \cite{Kyle-Ballantine:2019aa}}}
\author{Michael Mazilu\\
SUPA, School of Physics and Astronomy, \\
University of St Andrews, United Kingdom \\
E-mail: mm17@st-andrews.ac.uk}

\begin{document}
\maketitle

\begin{abstract}
Optical micro-manipulation and trapping of micro-particles delivers a mechanical system in direct interaction with a beam of light. In this interaction, the optical properties such as polarisation, beam profile and wavelength of the trapping beam are important. Different beams are associated with different momentum transfer, trap stiffness and stabilisation properties, for example. One method to determine the best beam profile is through the use of the optical eigenmode approach. To use this method, we employ Mie scattering theory which enables the exact determination of the scattered field as a function of the incident field. 
More precisely, this approach allows us to calculate the Hermitian  relationship between the incident field and the optical forces acting on the scattering objects. This Hermitian relationship defines also a set of orthogonal optical eigenmodes which deliver a natural basis to describe momentum transfer in light-matter interactions. This relationship defines also a set of orthogonal optical eigenmodes.  Using these modes it is possible to define, for each numerical aperture, particle size or geometry, the optimal trapping beam.
\end{abstract}

\section{Introduction}

The eigenmode representation of fields can be utilised to describe light propagating in optical system such as  photonic crystals and waveguides. The Optical Eigenmodes (OEi) method generalises this approach to include any interaction that has a  quadratic (Hermitian) form with respect to the electromagnetic (EM) fields~\cite{Mazilu:11,Mazilu:2009dy,Mazilu:2011gx}. Example of quantities having a quadratic form are energy density, momentum, and angular momentum of the EM fields and these quantities correspond to forces or torques experienced by the scattering objects. The OEi method determines field profiles that are orthogonal to each other with respect to the measured quantity. These fields contribute independently to the measure and do not interfere. This decomposition allows for example the coupling of light to individual plasmonic nano-antennas arranged in an array while minimising cross-talk~\cite{Kosmeier2013}. 
The approach defines a set of orthogonal eigenmodes whose eigenvalue correspond to the quantity measured. These eigenvalue can therefore be used to optimise the measure. Optimising quantities such as spot size\cite{Mazilu:2011uf,Baumgartl:2011bm} or energy transmission is as trivial as choosing the mode with the highest or lowest eigenvalue.
The determination of the OEi eigenmodes can be done experimentally, numerically and/or theoretically. 

Scattering is one of the simplest light mater interactions possible. For spherical particles, this process can be described using the Lorenz-Mie theory, which makes use of vector spherical harmonic solutions of Maxwell’s equations to represent the fields involved. Using these solutions it is possible to describe the light field scattered from microscopic spherical particles and thus represent the field around a scattering object as a function of the incident fields. These solutions also allow us to determine the optical momentum transfer to the scattering object. This can be calculated using Maxwell stress tensor. In this paper, we will study the momentum transfer to a spherical particle illuminated by a superposition of vector Bessel beams and introduce the use of optical eigemodes to describe the optical forces acting of the microsphere and how to use the OEi decomposition and beam symmetries to optimise the optical trapping strength for 3D Mie scattering particles generalising the approach taken in~\cite{Michael-Mazilu:2012aa}.

\section{Theory}
In the following, we work with monochromatic fields ($\omega=k_0 c$) which define the incident field as a superposition of Bessel beams that are either azimuthally (s-polarisation)  or radially polarised (p-polarisation). Each Bessel beam is further characterised by its transversal wavevector $k_t=n_0 k_0 \sin(\gamma)$ and longitudinal wavevector $k_z=n_0 k_0 \cos(\gamma)$ where $\gamma$ and $n_0$ are the cone half-angle of the Bessel beam and index of refraction of the propagation media, respectively. 

In a first step, we define the vector Bessel beams in Cartesian coordinate system and their radial component in the spherical coordinate system. This component allows us to define the beam shape coefficients that together with the momentum transfer operator can be used to calculate the optical forces acting on spherical particles. The eigenmodes of the momentum operator can then be introduced, reducing the dimensionally of the system when calculating the momentum transfer. 

\subsection{Vector Bessel beam shape coefficients}

To simplify the definition of the Bessel beam in Cartesian coordinates, we introduce the following function:
\begin{eqnarray*}
B_\ell(k_t \rho)&=&(i)^\ell J_\ell(k_t \rho)e^{i\ell\phi}
\end{eqnarray*}
with $\rho=\sqrt{x^2+y^2}$ and $\phi=\arctan(y/x)$ and where $J_\ell(r)$ are the Bessel functions of the first kind.

In this case, the s-polarised Bessel beam takes the form:
\begin{equation*}
^{}{\bf E}^{\ell(s)}=\left(
\begin{array}{l}
^{}E^{\ell(s)}_x\\
^{}E^{\ell(s)}_y\\
^{}E^{\ell(s)}_z\\
\end{array}
\right)
=  c_w \left(
\begin{array}{l}
- i B_{\ell-1}(k_t \rho)+ i B_{\ell+1}(k_t \rho)\\
B_{\ell-1}(k_t \rho)+  B_{\ell+1}(k_t \rho)\\
0\\
\end{array}
\right)
\end{equation*}

\begin{equation*}
^{}{\bf H}^{\ell(s)}=\left(
\begin{array}{l}
^{}H^{\ell(s)}_x\\
^{}H^{\ell(s)}_y\\
^{}H^{\ell(s)}_z\\
\end{array}
\right)
=   c_w \frac{n}{c\mu_0}  \cos(\gamma) \left(
\begin{array}{l}
- B_{\ell-1}(k_t \rho)- B_{\ell+1}(k_t \rho)\\
- i B_{\ell-1}(k_t \rho)+i B_{\ell+1}(k_t \rho)\\
2  B_{\ell}(k_t \rho) \tan(\gamma) \\
\end{array}
\right)
\end{equation*}
where the $c_w=\pi \sin(\gamma)\sqrt{\cos(\gamma)} \exp(-i\omega t +i k_z z)$.

The p-polarised beam can be related to the s-polarised one via: $^{}{\bf E}^{\ell(p)}=-c\mu_0/n ^{}{\bf H}^{\ell(s)}$ and $^{}{\bf H}^{\ell(p)}=n/(c\mu_0) ^{}{\bf E}^{\ell(s)}$. The main property of these Bessel beams is that they are eigenfunctions of the solid rotation operator with a integer valued eigenvalue $\ell$ 
\begin{eqnarray*}
-i{\bf e}_z\times{\bf E}^{\ell(s,p)}-i\left( {\bf r }\times\nabla\right)_{{\bf e}_z}{\bf E}^{\ell(s,p)}&=&\ell{\bf E}^{\ell(s,p)}\\
-i{\bf e}_z\times{\bf H}^{\ell(s,p)}-i\left( {\bf r }\times\nabla\right)_{{\bf e}_z}{\bf H}^{\ell(s,p)}&=&\ell{\bf H}^{\ell(s,p)}
\end{eqnarray*}
where ${\bf e}_z$ is the z-direction unit vector. The two terms in this operator can be identified with the spin angular momentum and the orbital angular momentum, respectively. 

Projecting the electric and magnetic fields on the radial unit vector in spherical coordinates defines the radial components of the electromagnetic fields as
\begin{eqnarray*}
^{}E^{\ell(s)}_r(r,\theta,\phi)&=&  i^\ell c_s  \exp(i\ell \phi)\left(
-J_{\ell-1}(k_t r \sin\theta)\sin\theta- J_{\ell+1}(k_t r \sin\theta) \sin\theta
\right) \\
^{}H^{\ell(s)}_r(r,\theta,\phi)&=&  i^\ell  c_s  \frac{n}{c\mu_0}  \exp(i\ell \phi)\\
&&(
 2J_{\ell}(k_t r \sin\theta) \cos\theta\sin\gamma+i J_{\ell-1}(k_t r \sin\theta) \sin\theta\cos\gamma  -iJ_{\ell+1}(k_t r \sin\theta) \sin\theta\cos\gamma)
\end{eqnarray*}
where
$c_s=\pi \sin(\gamma)\sqrt{\cos(\gamma)} \exp(-i\omega t +i k_z z \cos(\theta))$

The beam shape coefficients can then be determined by calculating the inner product of the radial field component with the spherical harmonic function $Y^m_n(\theta,\phi)$

\begin{equation*}
j_n(kr)\left(
\begin{array}{l}
-g^{TM}_{nm}\\
g^{TE}_{nm}/Z
\end{array}
\right)=\frac{kr}{\sqrt{n(n+1)}}
\int_0^\pi d\theta\int_0^{2 \pi}d\phi \;\;{Y^m_n}(\theta,\phi)^*
\left(
\begin{array}{l}
^{}E_r\\
^{}H_r
\end{array}
\right)
\end{equation*}
with $Z=\mu_0 c/n_0$ and where $*$ stands for the complex conjugate and where $g^{TM}_{nm}$ and $g^{TE}_{nm}$ are the beam shape coefficients. 

Introducing
$c_m=\pi \sin(\gamma)\sqrt{\cos(\gamma)} \exp(-i\omega t )$, we have for the s-polarised Bessel beam
\begin{eqnarray*}
g^{TM}_{nm}&=& \delta_m^\ell \frac{ 4 \pi i^{n} \sqrt{\pi\cos\gamma(2n+1)(n-m)!}}{\sqrt{ n(n+1)(n+m)!}} (m)P^m_n(\cos\gamma)\\
g^{TE}_{nm}&=&\delta_m^\ell
\frac{4 \pi  i^n \sqrt{\pi \cos\gamma(2n+1)(n-m)!}}{\sqrt{ n(n+1)(n+m)!}} (-i)
((n+1)\cos\gamma P^m_n(\cos\gamma)+(m-n-1)P^m_{n+1}(\cos\gamma))\\
\end{eqnarray*}
valid for $n\ge \ell$.
In the following, we define a single list of beam shape coefficients $$g_j=g^{p_j}_{n_jm_j}$$
where $m_j$, $n_j$ and $p_j$ are all coefficients considered indexed using the subscript $j$. The polarisation index $p_k$ can take the values of 1 and 2 corresponding respectively to $TM$ and $TE$ modes. A superposition of vector Bessel beams will then correspond to a superposition of beam shape coefficients $g_i$. 
Indeed, an angular dependent  superposition of Bessel beams defined this way makes it  possible to describe high-NA beams used in microscopes which introduce  spherical aberrations~\cite{Neves:2007p6055}.

\subsection{Momentum transfer}
Momentum transfer to scattering objects can be calculated using Maxwell's stress tensor defined by
\begin{eqnarray*}
\widetilde{\sigma} 
&=&\frac{1}{2} \left(\left(n_0^2\epsilon_{0}{\bf E}\cdot\mathbf{E}+\mu_{0}{\bf H}\cdot{\bf H}\right)\widetilde{I}-2 n_0^2\epsilon_{0}{\bf E}\otimes{\bf  E} -2\mu_{0}\mathbf{H}\otimes\mathbf{H}\right),
\end{eqnarray*}
 where $\otimes$ stands for the tensor product and $\widetilde{I}$
for the identity 3x3 matrix. 
  In a first instance, we are interested in the force in the $z$-direction which can be calculated by integrating over a spherical surface $S$ surrounding the object
\begin{equation}\label{integFz}
F_z=<\oint_S \mathbf{e}_z\widetilde{\sigma} d\mathbf{n} >
\end{equation} 
where $<>$ stands for the time average over the optical cycle. The fields $\mathbf{E}$ and $\mathbf{H}$ correspond to the sum of the incident and scattered fields. The incident fields are defined by the beam shape coefficients while the scattered fields can be calculated using the Mie scattering coefficients and the beam shape coefficients. Altogether, we remark that the optical force acting in the z-direction can be expressed in a quadratic form with respect to the beam shape coefficients.  This quadratic Hermitian form is based on the matrix operator
\begin{eqnarray*}
M_k^j=\frac{1}{2}
\delta_{m_k}^{m_j}\delta_{p_k}^{p_j} 
\left(\delta_{n_k}^{n_j+1}\sqrt{\frac{(1-n_k)^2(n_j^2-m_k^2)}{n_k^2(4n_k^2-1)}}-\delta_{n_k+1}^{n_j}\sqrt{\frac{(1-n_j)^2(n_k^2-m_k^2)}{n_j^2(4n_j^2-1)}}\right)\\
\left(\delta_{p_k}^1(b_k+b^*_j-2b_kb_j^*) +\delta_{p_k}^2 (a_k+a^*_j-2a_ka_j^*) \right) \\
+\frac{i}{2}\delta_{n_k}^{n_j} \delta_{m_k}^{m_j}
\frac{m_k}{n_k(n_k+1)}\left(\delta_{p_k}^{1} \delta_{2}^{p_j} (b_k+a_k^*-2b_ka_k^*)-\delta_{p_k}^{2} \delta_{1}^{p_j} (a_k+b_k^*-2a_kb_k^*) \right)
\end{eqnarray*}
where $n_k$ $m_k$ and $p_k$ is the $n$- and $m$-index and polarisation state of the $k$-th beam shape component and where $a_k$ and $b_k$ are the $n_k$-indexed Mie scattering coefficients of the scattering object.  

The optical momentum transfer in the z-direction is then defined by 
\begin{equation}\label{QF}
F_z=g^kM_k^jg_j
\end{equation}
where $g^k=g_k^*$. This quadratic Hermitian expression calculates the force acting in the z-direction on the scattering object positioned in at the origin of the coordinate system however the direction can be changed using spherical harmonics rotation and translation matrices.

\subsection{Optical eigenmodes}
The momentum transfer matrix is by construction Hermitian and in general all real field properties that can be expressed as a function of field in a quadratic form will lead to a matrix/operator that is Hermitian~\cite{Baumgartl:2011bm,DeLuca:2011jl}. The consequence of this observation is that the momentum transfer matrix defines a set of orthogonal vectors that correspond to the eigenvectors of $M_k^j$. Each of these eigenvectors defines a field that we call optical eigenmode (OEi) of the measure in questions, in this case it is the optical eigenmode of $F_z$. If the Mie scattering order tends to infinity then the optical eigenmodes will form a complete Hilbert basis set of solutions of Maxwell's equations. 
Each quadratic  measure will define an operator with its own set of optical eigenmodes. If two operators commute then it is possible to define a set of optical eigenmodes that are simultaneously eigenmodes for each measure (operator).  

The eigenvalues can also be used to define optimised beams with respect to a specific measure ie. the OEi of the $F_z$ measure having the largest eigenvalue corresponds to the beam delivering the largest momentum transfer to the particle at a constant incident power. Further, the eigenvalues can be used to sort the optical eigenmodes by order of importance  such that we can reduce the degrees of freedom for which a system needs to be solved for by discarding all optical eigenmodes that are not contributing. It is this property that can be used to improve the computational speed of the numerical model. We determine the optical eigenmodes that lead to any significant momentum transfer and use these modes to describe the incident field. Any incident field that does not couple to the optical eigenmodes will not lead to any measurable momentum transfer due to the symmetries of the problem considered. 

\subsection{Optimised trapping}

To define the best optimal trap, we restrict the illumination to cylindrical symmetric beams. These beams, due to their symmetry, will induce no transversal force on the particle. We then offset the particle and calculate the optical eigenmode with the largest eigenvalue corresponding to the largest restoring force. This delivers the optimal transversal trap for the spherical Mie particles. This approach generalises the numerical simulations optimising the 2D trapping of cylinders\cite{Michael-Mazilu:2012aa}.

\section{Numerical example}

The procedure outlined above can be used to calculate the optical trapping strength of trapped homogeneous or coated micro-particles including optical aberration introduced by high-NA microscope objectives\cite{Craig:2015kn}. 
Optical forces for larger particles can also be calculated and it is possible to determine the complex trajectories of levitated micro-particles in vacuum\cite{Mazilu:2016jn}.

\begin{figure}[htbp] 
   \centering
   \includegraphics[trim={0 0 0 2cm},clip,width=13cm]{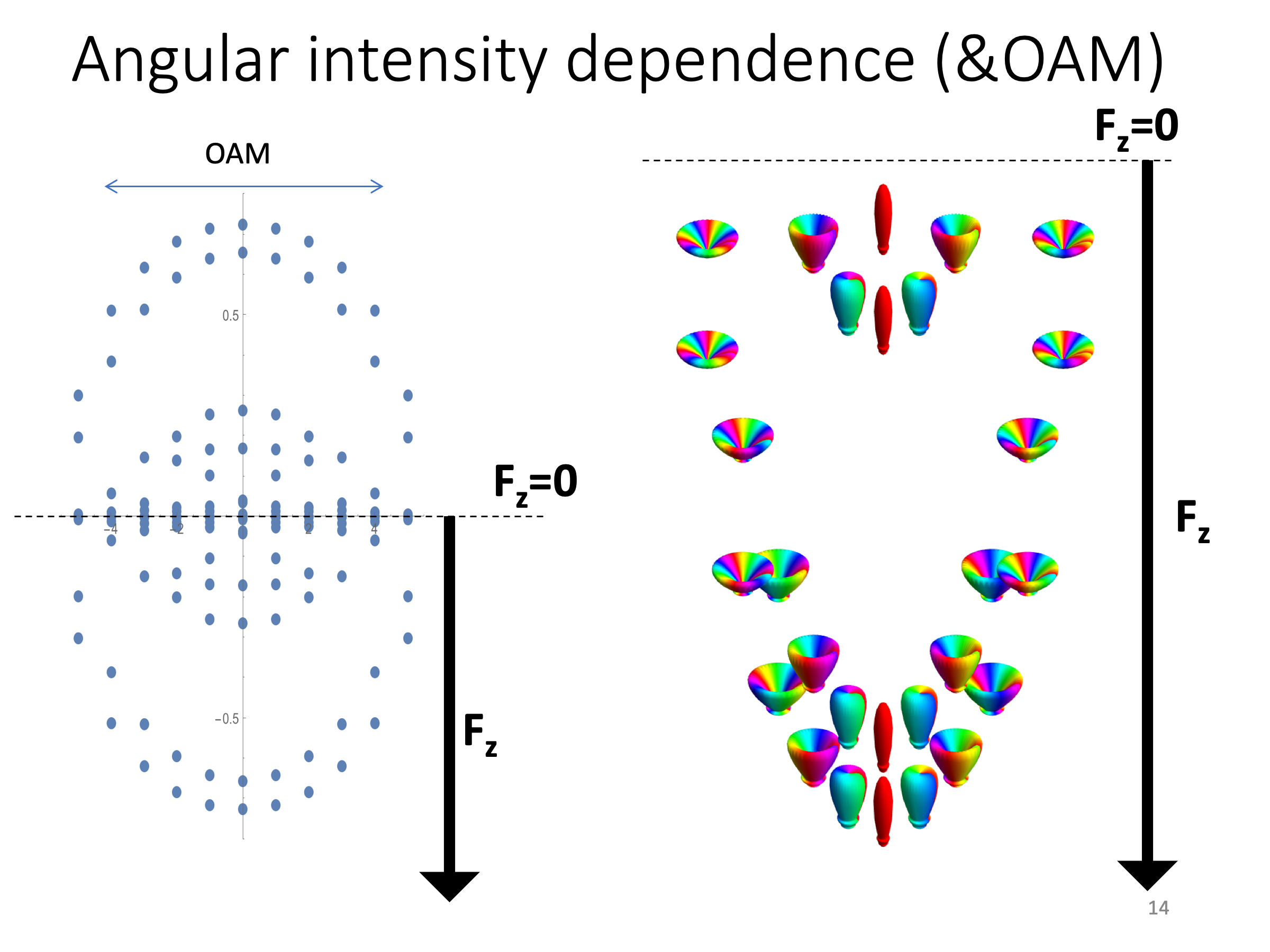}
   \caption{(a) Eigenvalue representation of the bound momentum (linear $F_z$ and orbital $L_z$) optical eigenmodes. (b) Angular intensity distribution of the optical eigenmodes. The colour indicate the phase of the field.}
   \label{fig1}
\end{figure}

\begin{figure}[htbp] 
   \centering
   \includegraphics[trim={0 0 0 2cm},clip,width=13cm]{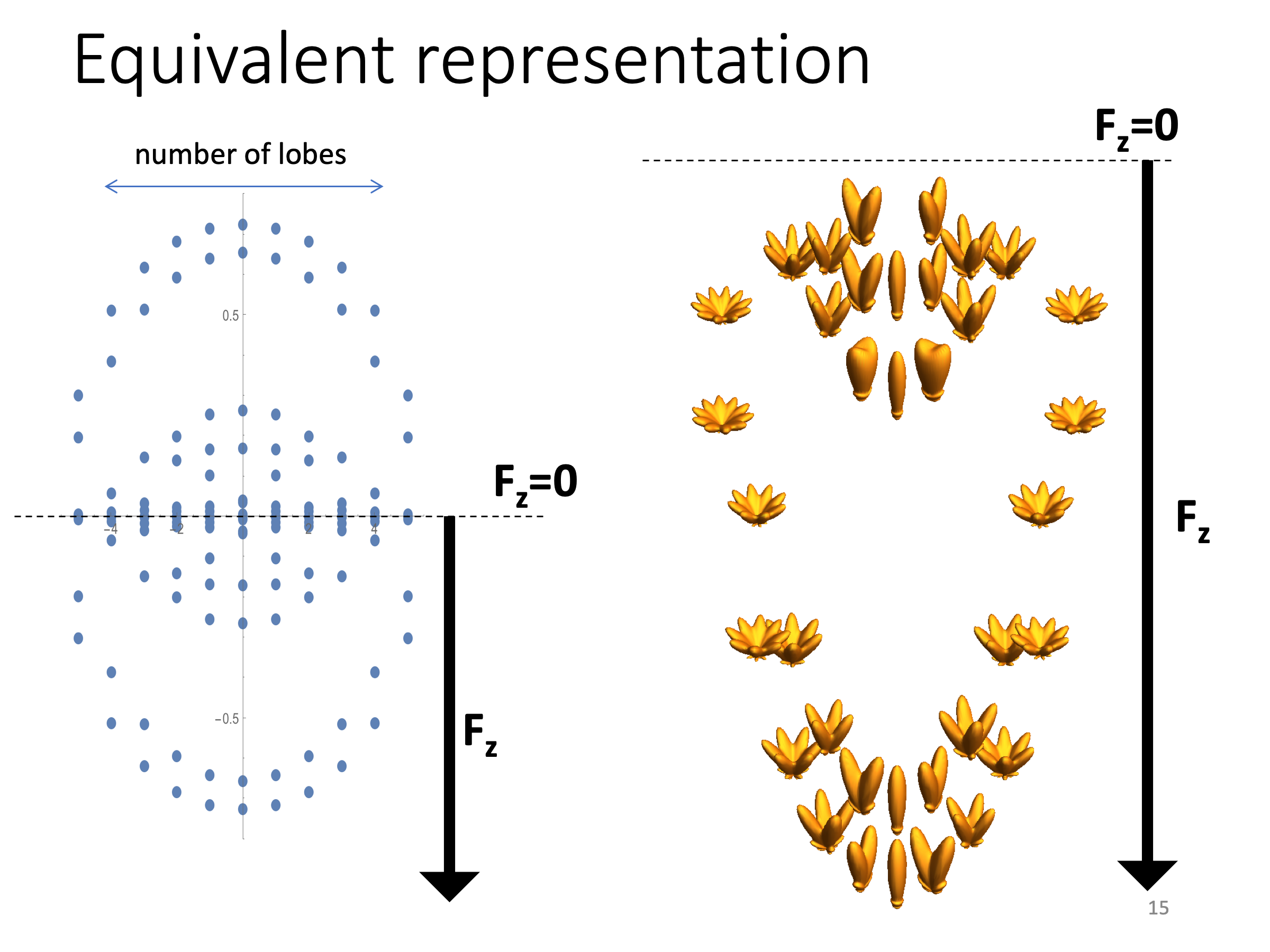}
   \caption{(a) Eigenvalue representation of the bound linear momentum, $F_z$,  optical eigenmodes. The horizontal axis counts the number of lobes and their position. (b) Angular intensity distribution of the optical eigenmodes. Note, the fields are real and there is no orbital angular momentum. These modes are obtained by hybridisation of the modes in Fig.\ref{fig1}. }
   \label{fig2}
\end{figure}

\begin{figure}[htbp] 
   \centering
   \includegraphics[width=13cm]{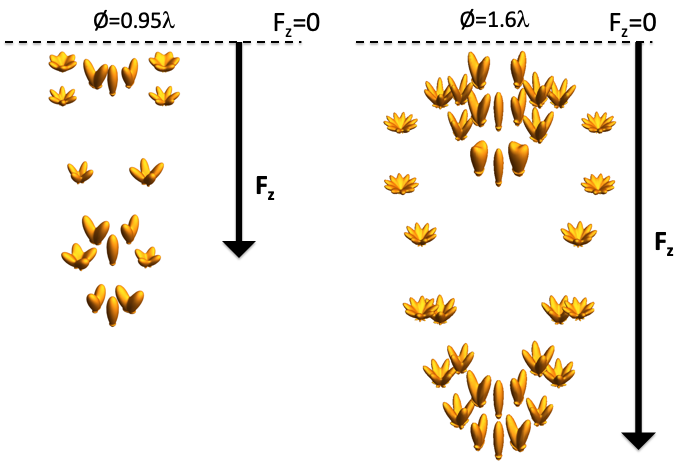}
   \caption{Angular intensity distribution of the optical eigenmodes for two particles of different sizes (index of refraction  1.5).}
   \label{fig3}
\end{figure}

\begin{figure}[htbp] 
   \centering
   \includegraphics[width=13cm]{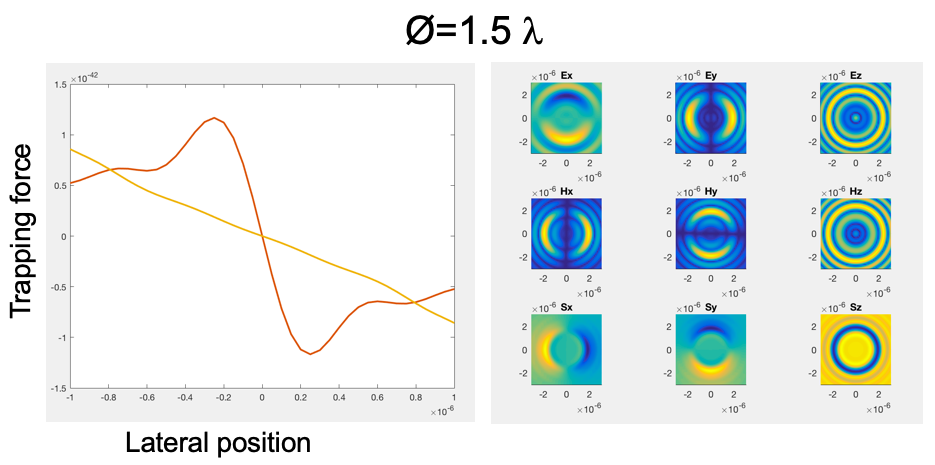}
   \caption{(left) Transverse trapping forces for a Gaussian beam in yellow and for the optical eigenmode optimal trapping beam in red (arbitrary units). (right) Electric field, magnetic field and Poynting vector components of the optical eigenmode in the trapping plane.  }
   \label{fig4}
\end{figure}

\begin{figure}[htbp] 
   \centering
   \includegraphics[width=7cm]{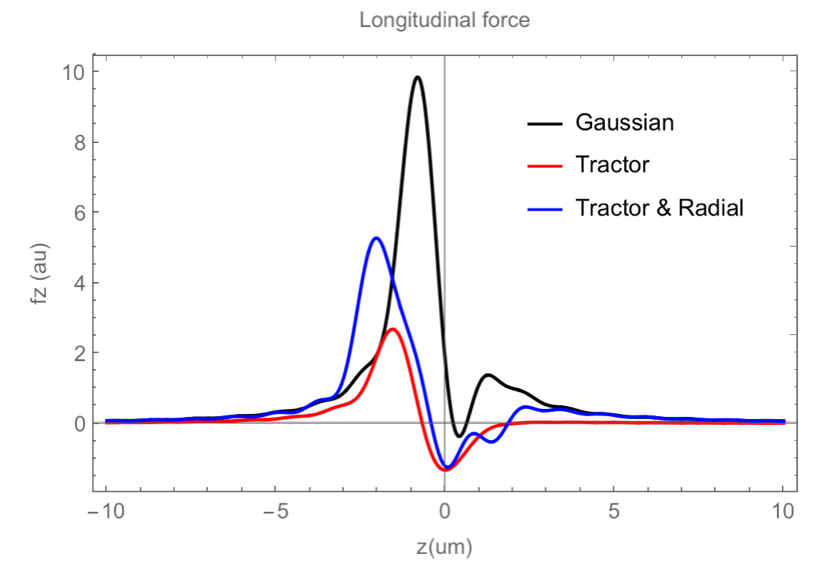}~\includegraphics[width=7cm]{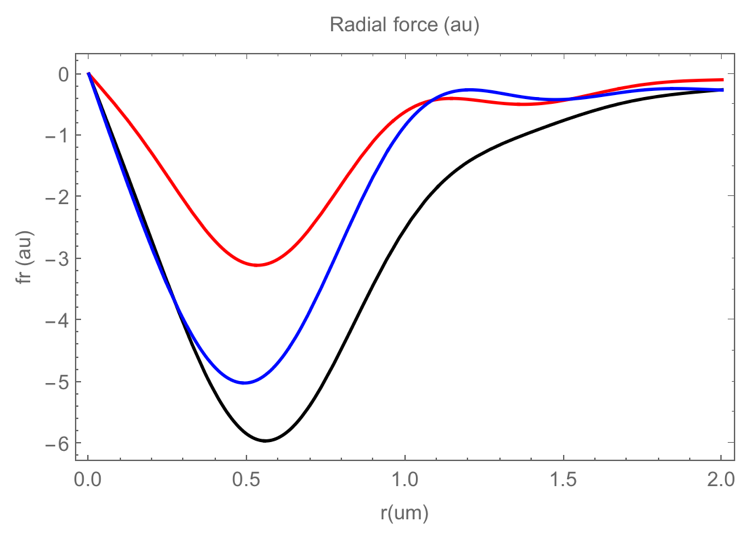}
   \caption{(left) Longitudinal trapping forces for a Gaussian beam in black and for the optical eigenmode optimising the ``tractor beam'' in red and  for an optimised superposition between ``tractor beam" and the radial trapping eigenmode. (right) Transverse forces for the same beams as in left panel. Note that the transversal trapping position is at $r=0$. }
   \label{fig5}
\end{figure}

\subsection{Optical eigenmodes and their particle size dependance}

Each optical eigenmodes has a real eigenvalue associated with it. This value corresponds in this case to the optical force acting in the z-direction on the scattering object when the incident field is the optical eigenmode. Considering the set of all operators that commute with each other then the set of  eigenvalues from these operators can be used to uniquely identify an optical field (see Figs. \ref{fig1}, \ref{fig2} and \ref{fig3}). The last figure shows that the number of optical eigenmodes increases with the size of the particle. This can be seen as an increase in the number of optical degrees of freedom of the optical system\cite{Chen:2016cp}, which in this case directly relates  to the number of whispering gallery modes resonant in the microsphere. This number increases as the particles size increases. We further remark that the angular structure of the momentum optical eigenmodes mimics the whispering gallery modes leading to an enhanced momentum transfer similar to the one observed in 2D~\cite{Michael-Mazilu:2012aa}.

\subsection{Optimal trapping beam}

As discussed in the theoretical section, we restrict the incident beam Hilbert space to cylindrical symmetric beams which in any superposition will always create a net zero transverse force on the particle. Within this restricted Hilbert space we use (\ref{QF}) in the transversal direction for a particle that is displaced with respect to the centre of the beam. The optical eigenmode with the largest momentum eigenvalue corresponds to the optimal transversal trapping beams for the spherical particle. In this case, we considered a particle with a diameter of 1.5$\lambda$ and index of refraction of 1.5. Figure (\ref{fig4}) shows a ten fold enhanced trap stiffness with respect to a Gaussian beam of the same power. 

Similar Hilbert space restriction be used in the case of 3D trapping in a $4\pi$ system where we can consider any beam and its counter propagating beam. However, the majority of trapping configurations allows only access to a restricted numerical aperture that is single sided. There are multiple options when considering three dimensional trapping in the case of single sided illumination. The simplest option is to search for the optimal beam profile delivering an optical force against the direction of the beam. This beam corresponds to the optimal ``tractor beam" and due to the interplay between scattering force and gradient force, this beam will always be accompanied by a stable longitudinal trapping position (see Figure \ref{fig5}). Further, the ``tractor beam'' part of the beam counteracts the effect of Brownian motion stabilising the longitudinal trapping. 

However, the Hermitian matrices corresponding to the transversal and longitudinal momentum transfer to the spherical particle do not commute with each other.  It is therefore not possible to optimise using the optical eigenmode procedure both trap properties ie. these properties are not independent of each other.  In this case, we have calculated the two optical eigenmodes (transversal and longitudinal) and optimised the phase difference between the two beams to achieve best transversal  trapping strength  (see Figure \ref{fig5}). 

\bibliographystyle{unsrt}
\bibliography{abs}

\end{document}